\documentstyle[11pt]{article}
\begin{document}
\noindent{\em Preprint  KONS-RGKU-99-01}\hfill{\em Submitted to
Phys.Rev. A}
\vskip1cm

\begin{center}
{\Large \bf
Aharonov-Bohm scattering of charged particles \\ and neutral atoms: the role of absorption }
\\[1cm]
{\bf  J\"urgen Audretsch$^{\dagger,}$\footnote{{\it e-mail:
Juergen.Audretsch@uni-konstanz.de}} and
Vladimir D. Skarzhinsky$^{\dagger, \ddagger,}$\footnote{{\it e-mail:
vdskarzh@sgi.lebedev.ru}}}
\\[0.5cm]
$^\dagger$Fakult\"at f\"ur Physik der Universit\"at Konstanz,
Postfach M 673,\\ D-78457 Konstanz, Germany \\[0.5cm]
$^\ddagger$P.~N.~Lebedev Physical Institute, Leninsky prospect 53,
Moscow 117924, Russia
\\[1cm]

\begin{abstract}
\noindent
The Aharonov-Bohm scattering of charged particles by the magnetic field of an infinitely long and infinitely thin solenoid (magnetic string) in an absorbing medium is studied. We discuss
the partial-wave approach to this problem and show that standard partial-wave method can be adjusted to this case. The effect of absorption leads to oscillations of the AB cross section.

Based on this we investigate the scattering of neutral atoms with induced electric dipole moments by a charge wire of finite radius which is placed in an uniform magnetic field. The physical realistic and practically important case that all atoms which collide with the wire are totally absorbed at its surface, is studied in detail. The dominating terms of the scattering amplitude are evaluated analytically for different physical constellations. The rest terms are written in a form suitable for a numerical computation. We show that if the magnetic field is absent, the absorbing charged wire causes oscillations of the cross section. In the presence of the magnetic field the cross section increases and the dominating Aharonov--Bohm peak appears in the forward direction, suppressing the oscillations.


\end{abstract}
\end{center}

\section{Introduction}

The investigation of the influence of magnetic fluxes on charged quantum systems was initiated by the famous paper of Aharonov and Bohm~\cite{Aharonov59}. It is worth to mention that the related effect has been discussed earlier in Refs.\cite{Franz39,Ehrenberg49}. The Aharonov-Bohm (AB) effect has been observed in numerous experiments based on electron interference (for reviews see \cite{Olariu85,Peshkin89}.)

Many papers are devoted to different aspects of the AB effect. For consequences in the domain of quantum electrodynamics see \cite{our papers}. The AB phase is a particular case of a geometrical phase \cite{Anandan97}. Aharonov and Casher \cite{Casher84} have pointed out a similar topological effect.

Wilkens reversed in \cite{Wilkens94} the Aharonov-Casher situation and derived the quantum phase acquired by an electric dipole moving in the radial field of a not very realistic straight line of Dirac magnetic monopoles. It were Wei, Han and Wei \cite{Wei95} who proposed a configuration which is experimentally feasible: The radial electric field of a charged wire polarizes scattered neutral atoms. A uniform magnetic field is applied parallel to the wire. The atoms moving in these two fields will then acquire a quantum phase and experience scattering of the AB type. The corresponding Lagrangian is formally equivalent to the Lagrangian of a charged particle moving in an effective AB potential, which is now proportional to the magnetic field itself, and in the potential of the electric field of the wire. The magnetic field yields no force on the particle, nevertheless a pure topological quantum phase is attached to the wave function. This phase should be measurable by means of atom interferometric techniques. The corresponding experiments are on their way. Refraction in the limiting case of the vanishing magnetic field (only radial symmetric electric field of the charged wire is present) has already been observed \cite{Pfau98}.

In this article we discuss the scattering of polarizable atoms in the magnetic and electric fields described above. According to the strength of the fields and the radius of the charged wire, different physical situations have to be distinguished. Our treatment of the respective scattering amplitudes and elastic cross sections is based on the partial-wave approach. A quick access to the results for a particular choice of the physical parameters has been sketched in a letter \cite{Audretsch98}.

An important practical aspect of the experimental set up is the fact that every wire has a finite radius. All atoms which fall on the wire loose their polarization and don't take part in the scattering process furthermore, which thereby becomes inelastic. The wire defines a radius of absorption $\rho_{\rm abs}$ where total absorption happens. The other constitutive effect goes back to the attraction of the wire. Incident atoms will fall on the wire if they don't have sufficient angular momentum to escape. This defines a critical radius $\rho_{\rm c}.$ Partial-wave analysis enables us to represent these two radii and to establish total absorption in a phenomenological way. As we understand it, the authors of Ref. \cite{Leonhardt98} discuss the case of a wire with exactly vanishing radius. This is not only unphysical but may imply severe mathematical problems too, as has been pointed out by Hagen \cite{Hagen96}.

From the mathematical point of view the difference between the scattering of polarized atoms described above and the AB scattering of charged particles is i) the presence of the additional radial attractive electric field and ii) the absorption on the surface of the wire. We approach these modifications of the set up step by step and discuss in Sec.2 the AB scattering with absorption first. This is a physical problem which is in itself of interest. To be able to introduce absorption in a phenomenological way, we give a partial-wave analysis of the AB scattering in Sec.2.2 where we reproduce the well known AB scattering amplitude (comp. Sec.2.1) in this way. We thereby describe the incident flux of quanta by a modified plane wave as it follows from the integral representation approach \cite{Aharonov59} to the AB scattering of Sec.2.1. The problems, which may occur in the phase shift method and which have been pointed out in Ref.\cite{Hagen90}, therefore don't occur.

Absorption is then introduced as a process which reduces the intensity of the outgoing flux of the elastic scattering. The influence of total absorption results in a modified scattering amplitude $f_{AB}^{\rm abs}(\varphi)$ which is derived and discussed in Sec.2.3.

Turning to the scattering of neutral atoms with induced electric dipole moments, we show in Sec.3.1 and 3.2 how the attraction by the charged wire causes a particular form of elastic scattering. Again we have to introduce in addition the absorption on the surface of the wire. Depending on whether the radius of absorption $\rho_{\rm abs}$ is smaller or larger as the critical radius $\rho_{\rm c}$ we work out in Sec.3.3 the result for the respective scattering amplitude $f^{\rm abs}(\varphi)$ in the form $f^{\rm abs}(\varphi)=f^{\rm abs}_{\rm AB}(\varphi)+f_w (\varphi).$ The influence of the charged wire is then localized in $f_w (\varphi).$ The influence of the strength of the magnetic field on $f^{\rm abs}(\varphi)$ and its behaviour for small scattering angles is discussed. The limiting case that there is only a charged wire and no magnetic field, which has already been treated in Ref.\cite{Denschlag97} is stated in Sec.3.4 and completes our investigation.


\section{Aharonov--Bohm scattering with absorption}

In this section we study the scattering of charged particles by the magnetic field of an infinitely long and infinitely thin solenoid (magnetic string) embedded in a medium which absorbs particles.

\subsection{Pure Aharonov--Bohm scattering of charged particles}

In a first step we consider the scattering of particles with mass $M$ and charge $e$ by a solenoid of infinite length and arbitrarily small radius (line flux) lying on the $z$-axis. We sketch first the calculations which go back to Aharonov and Bohm (AB) \cite{Aharonov59,Franz39,Ehrenberg49} (for the comprehensive review see \cite{Olariu85,Peshkin89}). Due to the symmetry of the background field the problem reduces to two dimensions. The vector potential may be written
\begin{equation}\label{A}
eA_\varphi = {\hbar c\beta\over\rho}, \quad \beta=\Phi /\Phi _{0}=
N+\delta, \quad 0\le\delta<1, \quad N=0,1,2,3,...\,.
\end{equation}
where $\Phi$ is the magnetic flux and $\Phi _{0} = 2\pi {\hbar}c/e$ is the magnetic flux quantum. $\rho$ and $\varphi$ are cylindrical coordinates.

The stationary Schr\"{o}dinger equation in the plane $z=0$ takes the form:
\begin{equation}\label{se}
\left\lbrace {\partial ^{2}\over \partial \rho ^{2}}
+ {1 \over \rho }{\partial \over \partial \rho } - {1\over\rho^2}
\left[ - i {\partial \over \partial  \varphi } - \beta \right] ^{2} + p^2\right\rbrace \psi(\rho, \varphi)  = 0
\end{equation}
where $p^2 = (2M/{\hbar}^2){\cal E}$ with ${\cal E}$ being the particle energy.

This equation can be solved for the scattering case using the partial-wave decomposition $(m=0, \pm 1, \pm 2, ...)$
\begin{equation}\label{psi}
\psi_{AB}(\rho ,\varphi ) = \sum_{m=-\infty}^{\infty}
R_m(\rho )\,e^{im \varphi}
\end{equation}
with radial functions which obey the radial equation
\begin{equation} \label{re}
R''_m + {1\over \rho } R'_m - {(m-\beta)^2\over \rho^2} R_m + p^2 R_m = 0.
\end{equation}
\noindent
We assume that the angle $\varphi_p$ under which the particles fall in, is equal to zero. For arbitrary $\varphi_p$ replace $\varphi$ by $\varphi-\varphi_p.$ The regular solutions of Eq.(\ref{re}) are
\begin{equation}\label{rs}
R_m(\rho) = c_m J_{|m - \beta |}(p \rho)
\end{equation}
with the coefficients
\begin{equation}\label{c}
c_m = e^{i\pi m - i{\pi \over 2}|m - \beta|}\,.
\end{equation}
$J_{|m-\beta|}(x)$ are the Bessel functions. The wave function $\psi_{AB}(\rho ,\varphi )$ is then obtained in the form
\begin{eqnarray}\label{swf}
\psi_{AB}(\rho ,\varphi ) &=& \sum_{m=-\infty}^{\infty}
e^{i\pi m - i {\pi \over 2} |m - \beta|}\,J_{|m - \beta |}(p\rho )\, e^{im \varphi} \nonumber \\
&=& e^{iN(\varphi-\pi)}\,\sum_{m=-\infty}^{\infty}
e^{i\pi m - i {\pi \over 2} |m - \delta|}\,J_{|m - \delta|}(p\rho )\, e^{im \varphi}\,.
\end{eqnarray}
This solution is invariant under the substitution $\beta\rightarrow -\beta,\, \varphi\rightarrow -\varphi.$ Therefore we may restrict to $\beta > 0, \;(N\geq 0)$

The series (\ref{swf}) converges uniformly at all values of $\varphi$ and $\rho$ apart from $\rho\rightarrow\infty.$ Therefore it is not possible to study the asymptotic behaviour of $\psi_{AB}(\rho,\varphi )$ in starting from the expression (\ref{swf}) and using the asymptotic behaviour of the Bessel functions. This would require that $p\rho$ has to be large as compared to the order $|m-\delta|$ of the Bessel functions. It had been pointed out in Ref.\cite{Hagen90} that for large $|m-\delta|$ values in the summation the use of the asymptotic form becomes inappropriate. One has to sum up first and then go to the limit $\rho\rightarrow\infty.$ We will show below how nevertheless the partial analysis of the scattering problem can be adapted to the AB scattering problem taking explicitly into account that the vector potential (\ref{A}) decreases slowly at infinity.

The traditional approach to obtain the asymptotic behaviour from Eq.(\ref{swf}), which goes also back to the pioneer paper \cite{Aharonov59}, is to rewrite the r.h.s. as an integral:
\begin{eqnarray}\label{ifswf}
\psi_{AB}(\rho, \varphi ) &=& e^{iN (\varphi-\pi)}\,
e^{ip \rho \cos \varphi}\,{1\over 2} \sin\pi\delta \int_0^{p \rho} dx e^{-ix \cos \varphi} \nonumber \\
&&\lbrace e^{i{\pi \over 2}(1 - \delta )} H^{(1)}_{1-\delta }(x)
- e^{i {\pi \over 2} \delta} H^{(1)}_\delta (x)\,e^{i\varphi}
\rbrace
\end{eqnarray}
with the Hankel functions $H^{(1)}_\nu (x).$ For all directions apart from the direction $\varphi=0$ of the incident wave, the wave function $\psi_{AB}(\rho, \varphi)$ may then asymptotically $(\rho\rightarrow\infty)$ be separated into a  plane wave part $\psi^{\rm in}_{AB}(\rho, \varphi)$ describing the ingoing particles plus a scattered wave:
\begin{equation}\label{as}
\psi_{AB}(\rho,\varphi ) \rightarrow \psi^{\rm in}_{AB}(\rho, \varphi) + f_{AB}(\varphi)\, {e^{ip\rho}\over\sqrt{\rho }}\,.
\end{equation}
Because the AB potential of a straight flux line is a long-range potential, it has to be taken into account in the wave function $\psi^{\rm in}_{AB}(\rho, \varphi)$ of the incoming flux which will depend on $\beta.$ It takes the form of a modified plane wave
\begin{equation}\label{modpw}
\psi^{\rm in}_{AB}(\rho, \varphi) = e^{i\beta (\varphi-\pi)}\, e^{ip\rho\cos\varphi}\,.
\end{equation}
The scattering amplitude $f_{AB}(\varphi)$ turns then out to be for $\varphi \neq 0$
\begin{equation}\label{sa}
f_{AB}(\varphi )= - {1\over \sqrt{2\pi p}}\,e^{i N (\varphi-\pi)}\, e^{-i {\pi \over
 4}}\;e^{i {\varphi\over 2}}\; {\sin\pi\delta\over\sin{\varphi\over 2}}.
\end{equation}


\subsection{Partial-wave analysis}

In the following we restrict to the asymptotic region $\rho\rightarrow\infty$ and go back to the Eqs.(\ref{psi}) and (\ref{re}). The behaviour of the radial functions $R_m(\rho)$ is of the structure
\begin{equation}\label{asrs}
R_m(\rho)\rightarrow {1\over\sqrt{2\pi p\rho}}\left[e^{-i(p\rho-\pi m-{\pi\over 4})}+S_m\,e^{i(p\rho-{\pi\over 4})}\right]\,, \quad S_m = e^{2i\delta_m}\,.
\end{equation}
where the outgoing part contains the phase factor $S_m.$ For the solution (\ref{rs}) with the coefficients (\ref{c}) we get the phase shifts $\delta_m$:
\begin{equation}\label{phs}
\delta_m = {\pi\over 2}(|m|-|m-\beta|)\,.
\end{equation}

To take the incident wave $\psi^{\rm in}_{AB}(\rho, \varphi)$ explicitly into account in the asymptotic region, we expand it according to
\begin{equation}\label{dpw}
\psi^{\rm in}_{AB}(\rho, \varphi) = \sum_{m=\infty}^{\infty}
R^{\rm in}_m(\rho)\,e^{im\varphi}
\end{equation}
with radial functions
\begin{equation}\label{rdpw}
R^{\rm in}_m(\rho) = {1\over 2\pi} \int_0^{2\pi} d\varphi e^{i\beta (\varphi-\pi)}\, e^{ip\rho\cos\varphi}\,e^{-im\varphi}\,.
\end{equation}
Their asymptotic behaviour can be found using the stationary phase method:
\begin{equation}\label{ardpw}
R^{\rm in}_m(\rho) \rightarrow {1\over\sqrt{2\pi p\rho}}\left[e^{-i(p\rho-\pi m-{\pi\over 4})}+\cos\pi\beta\,e^{i(p\rho-{\pi\over 4})}\right]\,.
\end{equation}
We return to Eq.(\ref{asrs}) and rewrite $R_m(\rho)$ for $\rho\rightarrow\infty$ in the form
\begin{equation}\label{dasrs}
R_m(\rho)\rightarrow R^{\rm in}_m(\rho) + {1\over\sqrt{2\pi p\rho}}(S_m-\cos\pi\beta)\, e^{i(p\rho-{\pi\over 4})}\,.
\end{equation}
Having separated the term $R^{\rm in}_m(\rho)$ in Eq.(\ref{asrs}) we obtain the partial-wave amplitudes
\begin{equation}\label{fm}
f_m = {e^{-i{\pi\over 4}}\over\sqrt{p}}\left(S_m - \cos\pi\beta\right) = {e^{-i{\pi\over 4}}\over\sqrt{p}}\left(e^{i\pi(m-|m-\beta|)} - \cos\pi\beta\right)\,.
\end{equation}
Note that $f_m$ contains the characteristic term $\cos\pi\beta$ which will appear throughout the calculation below (see also Ref.\cite{Hagen90}).

The crucial test for the validity of this partial-wave approach is the agreement of the resulting scattering amplitude obtained by the partial decomposition
\begin{equation}\label{fab}
f(\varphi) = {1\over\sqrt{2\pi}}\sum_{m=-\infty}^{\infty} \;f_m\,e^{im\varphi}\,.
\end{equation}
with the AB scattering amplitude (\ref{sa}). To prove this and to show how Eq.(\ref{fab}) has to be handled, we rewrite $f(\varphi)$ according to
\begin{eqnarray}\label{f}
f(\varphi) &=& {e^{-i{\pi\over 4}}\over\sqrt{2\pi p}} \sum_{m=-\infty}^{\infty}
\left[ e^{i\pi(m-|m-\beta|}-\cos\pi\beta\right] \,e^{im\varphi}  \\
&=& {e^{-i{\pi\over 4}}\over\sqrt{2\pi p}}\, i\sin\pi\beta\left[\sum_{m=N+1}^{\infty} e^{im\varphi}\right.-
\left.\sum_{m=-\infty}^{N} e^{im\varphi} \right] \nonumber \\
&=& {e^{-i{\pi\over 4}}\over\sqrt{2\pi p}}\,e^{iN\varphi} i\sin\pi\beta\left[\sum_{m=1}^{\infty} e^{im\varphi}\right.-
\left.\sum_{m=-\infty}^{0} e^{im\varphi} \right]\nonumber \\
&=& {e^{-i{\pi\over 4}}\over\sqrt{2\pi p}}\,e^{iN\varphi} i \sin\pi\beta\left[\Sigma_+\varphi - \Sigma_-\varphi -1\right] \nonumber
\end{eqnarray}
where $\Sigma_\pm (\varphi)=\sum_{m=0}^{\infty} e^{\pm im\varphi}\,.$
As usual these sums can be evaluated and thus be given a meaning in replacing $\varphi\rightarrow\varphi\pm i\epsilon,\;{\rm with}\;\epsilon>0$ and taking the limit $\epsilon\rightarrow 0,$
\begin{eqnarray}\label{s+-}
\Sigma_\pm (\varphi)&=&\lim_{\epsilon\rightarrow 0} \sum_{m=0}^{\infty} e^{\pm im\varphi-m\epsilon} = \lim_{\epsilon\rightarrow 0}{1\over 1-e^{\pm i\varphi-\epsilon}} \nonumber \\
&=& P({1\over 1-e^{\pm i\varphi}}) + \pi \delta(1-\cos\varphi)\,.
\end{eqnarray}
Because we are physically interested again only in scattered particles for which $\varphi\neq 0,$ we can replace $P(...)$ by its argument. We then obtain from Eq.(\ref{f}) with Eq.(\ref{s+-})
\begin{eqnarray}\label{f'}
f(\varphi)&=& - {e^{-i{\pi\over 4}}\over\sqrt{2\pi p}} \, e^{iN\varphi} \,e^{i{\varphi\over 2}}{\sin\pi\beta \over \sin{\varphi\over 2}} \\
&=& - {e^{-i{\pi\over 4}}\over\sqrt{2\pi p}}\,e^{iN(\varphi-\pi)} \,e^{i{\varphi\over 2}}{\sin\pi\delta \over \sin{\varphi\over 2}} = f_{AB}(\varphi) \nonumber\,.
\end{eqnarray}
This completes the proof.


\subsection{The influence of absorption}

In the following we want to modify the physical situation discussed above in assuming that there is some cylindrically symmetric medium present which absorbs charged particles with an intensity which may depend on the radius $\rho.$ In this case elastic and inelastic scattering (absorption) happen in the same time. Following the standard procedure of the theory of inelastic scattering, we introduce absorption phenomenologically in the asymptotic region with reference to the partial-wave decomposition. This results in a modified wave function $\psi^{\rm abs}_{AB}(\rho, \varphi)$ which refers now only to the remaining (elastically) scattered particles. $\psi^{\rm abs}_{AB}(\rho, \varphi)$ will still obey the scattering condition (\ref{as}) with a new scattering amplitude $f^{\rm abs}_{AB}(\varphi)$ but of cause it does not correspond to a solution of Eq.(\ref{se}) furthermore because the physical situation has changed.

Absorption, like all other types of inelastic interaction, is here understood as a process which reduces the intensity of the elastic scattering. It is therefore most appropriately introduced with respect to the total partial flux per length through a fictitious cylinder of sufficiently large radius $\rho_0.$ We obtain with Eq.(\ref{asrs}) and for the limit $\rho_0\rightarrow\infty$
\begin{eqnarray}\label{pflux}
j^{\rm tot}_m(\rho_0) &=& \int_0^{2\pi}\rho_0\,d\varphi\,j_m (\rho_0, \varphi) \\
&=& \int_0^{2\pi}\rho_0\,d\varphi {\hbar\over 2iM} (R_m^* R'_m - R_m R'^*_m)|_{\rho=\rho_0}
\rightarrow - {\hbar\over M} (1-|S_m|^2)\,.\nonumber
\end{eqnarray}

If there is no absorption, we have $|S_m|=1$ so that the converging and diverging fluxes of $m$th partial-wave are equal (comp. Eq.(\ref{asrs})) and $j^{\rm tot}_m(\rho_0)$ vanishes. If absorption exists along with elastic scattering the converging flux of the elastically scattered particles is larger than the diverging flux so that the total partial flux is directed inwards: $j^{\rm tot}_m(\rho_0)<0.$ This corresponds to $|S_m|<1.$ Absorption can be total $|S_m|=0$ or weak $|S_m|\approx 1.$ Accordingly the absorption coefficient $\xi_m=1-|S_m|^2$ is equal to 1 or closed to zero. We mention that the total inelastic scattering cross section $\sigma^{\rm inel}$ is related to the total flux $J$ and to $|S_m|$ according to
\begin{equation}\label{flux}
J = \int_0^{2\pi}\rho_0\,d\varphi \sum_m j_m (\rho_0, \varphi) \rightarrow - {\hbar\over M}\sum_m (1-|S_m|^2) = - {\hbar p\over M} \sigma^{\rm inel}\,.
\end{equation}

Now we apply this scheme to the AB scattering and assume that total absorption happens for partial-waves with parameter $m$ out of the interval $[-n_-, n_+]:$
\begin{equation}\label{s=0}
S_m = 0 \quad \mbox{for} \quad -n_- \leq m \leq n_+\,.
\end{equation}
The corresponding partial scattering amplitudes are with Eq.(\ref{fm}):
\begin{equation}\label{mpw}
f_m=-{e^{-i{\pi\over 4}}\over\sqrt{p}}\,\cos\pi\beta\,.
\end{equation}

They do not vanish and contribute to the differential elastic scattering cross section. The asymptotic behaviour remains otherwise unchanged. We still have the scattering structure of Eq.(\ref{as})
\begin{equation}\label{asnew}
\psi_{AB}^{\rm abs}(\rho,\varphi ) \rightarrow \psi^{\rm in}_{AB}(\rho, \varphi) + f_{AB}^{\rm abs}(\varphi)\, {e^{ip\rho}\over\sqrt{\rho }}\,.
\end{equation}
where $f_{AB}^{\rm abs}(\varphi)$ is equal to $f(\varphi)$ of Eq.(\ref{fab}) with $f_m$ of Eq.(\ref{mpw}) if $m \in [-n_-, n_+]$ and with $f_m$ of Eq.(\ref{fm}) otherwise. The total flux $J$ is proportional to the total numbers $n^{\rm abs}$ of the modes with absorption
\begin{equation}\label{j}
J  = - {\hbar\over M}\,n^{\rm abs}\,.
\end{equation}
with
\begin{equation}\label{nabs}
n^{\rm abs}:=n_++n_-+1
\end{equation}
and is pointing redially inwards.

The interval $[-n_-, n_+]$ is fixed by the experimental situation which is to be described. If there is a totally absorbing cylinder of {\sl absorption radius} $\rho_{\rm abs}\neq 0,$ incident particles with impact parameter $a = |m-\beta|/p < \rho_{\rm abs}$ which refers to the kinetic angular momentum,  will be absorbed. This leads to
\begin{equation}\label{n+-}
n_\pm = [p\rho_{\rm abs} \pm \beta]\,.
\end{equation}
$[a]$ denotes the integer part of a quantity $a=[a]+\{a\}$ and is defined such that its fractional part \{a\} is positive. For $p\rho_{\rm abs}<\delta$ (including the limiting case $\rho_{\rm abs}=0$) we have $n_+=N,\; n_-=-N-1.$
\noindent

The elastic AB scattering amplitude in the presence of absorption can be written in the form ($\varphi\neq 0$)
\begin{eqnarray}\label{modsa}
&&f_{AB}^{\rm abs}(\varphi) = {e^{-i{\pi\over 4}}\over\sqrt{2\pi p}}
\left[\sum_{m=-\infty}^{\infty}\left(e^{i\pi (m-|m-\beta|)} -\cos\pi\beta\right)\, e^{im\varphi}\right. \nonumber \\
&-& \sum_{m=-n_-}^{n_+}\left(e^{i\pi (m-|m-\beta|)} -\cos\pi\beta\right)\, e^{im\varphi} + \left.\sum_{m=-n_-}^{n_+} \left(0-\cos\pi\beta\right) e^{im\varphi}\right] \nonumber \\
&=& f_{AB}(\varphi) - {e^{-i{\pi\over 4}}\over\sqrt{2\pi p}}
\sum_{m=-n_-}^{n_+} e^{i\pi (m-|m-\beta|)}\, e^{im\varphi}\,.
\end{eqnarray}
It turns out to be the AB amplitude $f_{AB}(\varphi)$ modified by a finite correction term. The effect of this correction term can be made explicit in rewriting $f_{AB}^{\rm abs}(\varphi)$ according to (comp. \cite{Audretsch98})
\begin{equation}\label{modsa1}
f_{AB}^{\rm abs}(\varphi) = - {e^{-i{\pi\over 4}}\over\sqrt{2\pi p}}\,
e^{i{n_+-n_-\over 2}\varphi}\,{1\over \sin {\varphi\over 2}}\,\sin\left(\pi\beta+{n^{\rm abs}\varphi\over 2} \right).
\end{equation}
If there is no absorption $(n^{\rm abs}=0)$ Eq.(\ref{modsa1}) coicides with Eq.(\ref{sa}).

\begin{figure}[h]
\let\picnaturalsize=N
\def\picsize{2.5in}
\ifx\nopictures Y\else{\ifx\epsfloaded Y\else\input epsf \fi
\let\epsfloaded=Y
{\hspace*{\fill}
\parbox{2.5in}{\ifx\picnaturalsize N\epsfxsize \picsize\fi \epsfbox{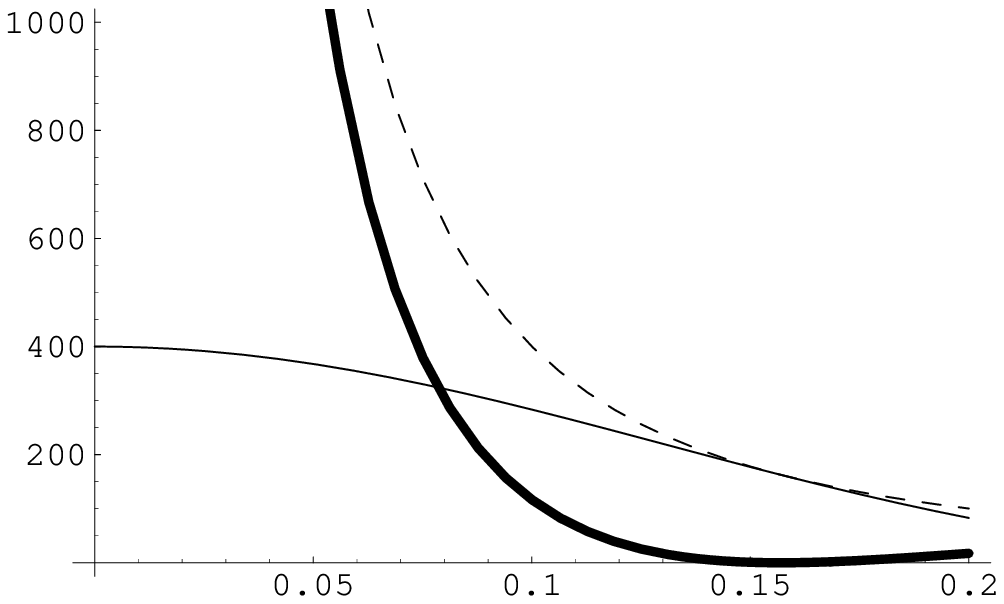}}\hfill
x\put(2,-46){{\small $\varphi$}}
\put(-165,60){{\small $\eta$}}
\put(200,-46){{\small $\varphi$}}
\put(27,62){{\small $\eta$}}
\put(-90,10){{\tiny $\leftarrow\beta=0.5, n^{\rm abs}=0$}}
\put(-150,10){{\tiny $\beta=0.5 \rightarrow$}}
\put(-150,-15){{\tiny $\beta=0 \uparrow$}}
\hspace{0.5cm}
\parbox{2.5in}{\ifx\picnaturalsize N\epsfxsize \picsize\fi \epsfbox{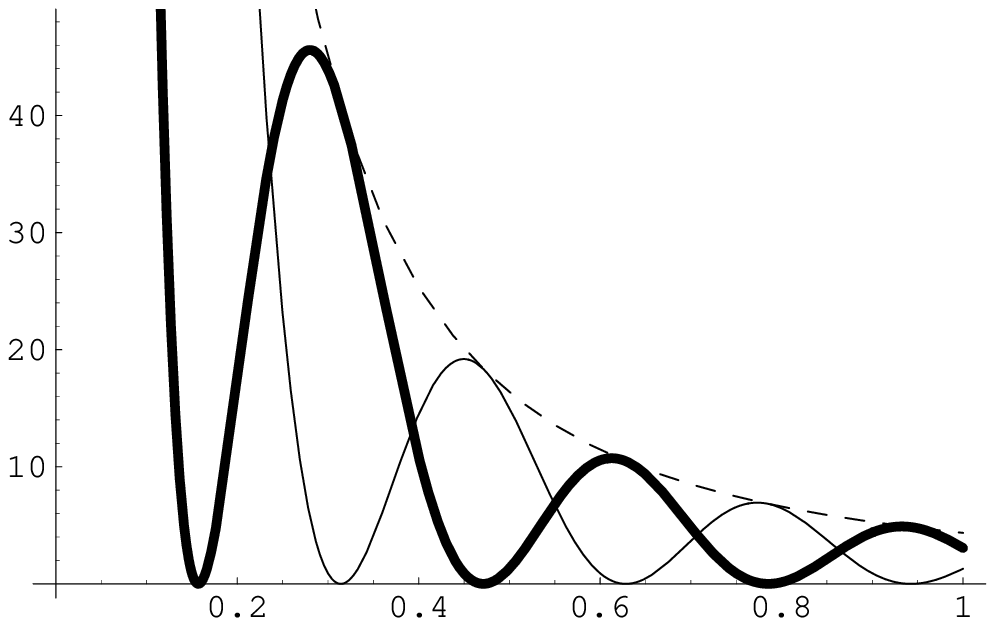}}\hspace*{\fill}}}\fi\\
\caption{Dimensionless differential cross section $\eta=2\pi p|f_{AB}^{\rm abs}(\varphi)|^2$ for elastic scattering by the magnetic flux $\beta=0.5$ with total absorption of 20 modes ($n^{\rm abs}=20$) as function of the scattering angle $\varphi$ (bold line). The special cases of vanishing magnetic field (thin line) or absence of absorption (pure AB scattering, dashed line) are shown as well.}
\end{figure}

The obtained formula shows the effects of total absorption on the one hand and presence of a magnetic field on the other (compare Fig.1). If there is only absorption (characterized by the number $n^{\rm abs}$ and $\beta=0$) the amplitude of the accompanying elastic scattering has the oscillatory behaviour $\sin {n^{\rm abs}\varphi\over 2}/\sin{\varphi\over 2}.$ Oscillations are typical manifestations of the presence of absorption. If in contrast to this there is no absorption we have the pure AB scattering. If both influences are present, we obtain an oscillatory behaviour for large scattering angles and the AB peak for small angles. Correspondingly, the differential cross section has dips and peaks of the width $\approx 2\pi/n^{\rm abs}$ located in points $\varphi=2\pi(k-\beta)/n^{\rm abs}$ and $\varphi=2\pi(k-\beta+1/2)/n^{\rm abs}, \; k=0, \pm1,...\,.$ The scattering pattern depends periodically on the intensity of the magnetic field.


\section{Aharonov--Bohm scattering of neutral atoms}


\subsection{The scattering amplitude}

We turn now to the scattering of polarizable atoms of mass $M_0$ and charge $e$ by the electric field of a straight charged wire which is collinear to a uniform magnetic field $B$ in $z$-direction. The axes of the wire is at $\rho=0.$ In a first step we assume a vanishing wire radius. Below we will show that for physical reasons this radius must be non-vanishing. We denote the electric polarizability by $\alpha$ and the homogeneous charge density on the wire by $\kappa$. The effective Lagrangian is then given by
\begin{equation}\label{L}
L = {1\over 2}Mv^2 + \alpha \,[\vec{B}\times\vec{E}]{\vec{v}\over c} +
{1\over 2}\alpha E^2
\end{equation}
in which we have omitted the terms of order $v^2/c^2.$ The important observation is that $L$ is equivalent to the Lagrangian of a particle with the same charge but mass $M=M_0+\alpha B^2/c^2$ moving in the effective AB potential
$e\vec{A}(\rho) = (\hbar c\beta/\rho) \vec{e}_{\varphi}$ (comp. (\ref{A})) and in the electric scalar potential $U_E(\rho) = - \alpha\kappa^2/(8\pi^2\rho^2).$ The magnetic field parameter $\beta$ is thereby $\beta := \alpha \kappa B/(2\pi\hbar c).$ This is reflected by the corresponding stationary Schr\"{o}dinger equation
\begin{equation}\label{se'}
\left\{{\hbar^2\over 2M}\left[{\partial^2\over\partial\rho^2} +
{1\over\rho}{\partial\over\partial\rho} -
{1\over\rho^2}\left(-i{\partial\over \partial\varphi} -
\beta\right)^2\right]  + {\alpha\kappa^2\over 8\pi^2\rho^2} +
{\cal E}\right\} \psi(\rho,\varphi) = 0
\end{equation}
with particle energy ${\cal E} = \hbar^2 p^2/2(M_0+\alpha B^2/c^2).$ Since even for atoms with large polarizability and for strong magnetic fields the magnetic energy will be smaller than the atom rest energy ($\alpha B^2/c^2\ll M_0 c^2$) we can omit the second term in $M$. Changing the value of $\beta,$ which in this case goes back to a changing magnetic field $B,$ causes a phase shift.

As in section 1.1 we solve this equation by means of a partial decomposition and obtain in analogy to (\ref{swf}):
\begin{equation}\label{swf'}
\psi(\rho , \varphi) = \sum_{m=-\infty}^{\infty} e^{i\pi m -
i{\pi\over 2}\nu}\; J_{\nu}(p\rho)\,e^{im\varphi}
\end{equation}
where
\begin{equation} \label{nu}
\nu^2 := m^2 - 2m\beta -\gamma^2, \quad \gamma^2 :=
{\alpha\kappa^2\over (2\pi\hbar)^2}M_0\,.
\end{equation}
The factor $e^{iN\varphi}$ could  be separated if the magnetic mass $\alpha B^2/c^2$ is neglected.

The series (\ref{swf'}) converges uniformly for all values of $\rho$ and
$\varphi$ except asymptotically for $\rho\rightarrow\infty.$ This is the same problem as the one we have had with series (\ref{swf}). In fact the series (\ref{swf}) is contained in (\ref{swf'}) because the parameter $\nu$ reduces asymptotically for large values of $m$ to the AB parameter $|m-\beta|.$ It is therefore advisable to subtract for $\varphi\neq 0$ the AB scattering wave function $\psi_{AB}(\rho ,\varphi )$ of Eq.(\ref{swf}) from $\psi(\rho ,\varphi )$ of Eq.(\ref{swf'}) to obtain
\begin{equation} \label{swf''}
\psi(\rho ,\varphi ) = \psi_{AB}(\rho ,\varphi ) + \delta\psi(\rho ,\varphi )
\end{equation}
with
\begin{equation}\label{dswf}
\delta \psi(\rho , \varphi) = \sum_{m=-\infty}^{\infty}\left[e^{i\pi m -
i{\pi\over 2}\nu}\;J_{\nu}(p\rho) - e^{i\pi m - i{\pi\over 2}|m-\beta|}\; J_{|m-\beta|}(p\rho)\right]\;e^{im\varphi}\,.
\end{equation}
This difference converges uniformly for all values of $\rho$ and $\varphi$ and contributes asymptotically only to the scattered flux
\begin{equation}\label{dswf'}
\delta \psi(\rho, \varphi) \rightarrow {e^{-i{\pi\over 4}}\over\sqrt{2\pi p \rho}}\,e^{ip\rho}\sum_{m=-\infty}^{\infty}\left(e^{i\pi m - i{\pi\over 2}\nu} - e^{i\pi m - i{\pi\over 2}|m-\beta|} \right).
\end{equation}
The ingoing wave $\psi_{AB}^{\rm in}(\rho, \varphi)$ remains unchanged, so that the asymptotic behaviour of the scattering wave function (\ref{swf'}) is again
\begin{equation} \label{as'}
\psi(\rho,\varphi)\rightarrow \psi^{\rm in}_{AB}(\rho, \varphi) + f(\varphi)\; {e^{ip\rho}\over
\sqrt{\rho}}\,.
\end{equation}
Comparison of Eq.(\ref{dswf'}) with Eq.(\ref{fm}) gives the scattering amplitude
\begin{equation}\label{psa'}
f(\varphi) = {1\over\sqrt{2\pi}}\sum_{m=-\infty}^{\infty} \;f_m\,e^{im\varphi} \,, \quad f_m = {e^{-i{\pi\over 4}}\over\sqrt{p}}\left(S_m - \cos\pi\beta\right)
\end{equation}
with the phase coefficients
\begin{equation}\label{phs'}
S_m = e^{2i\delta_m}, \quad \delta_m = {\pi\over 2}(m-\nu)\,.
\end{equation}

Important is that the parameter $\nu^2$ of Eq.(\ref{nu}) can take negative values for any values of the parameters $\beta$ and $\gamma.$ This happens when $m$ is inside a closed interval $[-m_-, m_+]$
\begin{equation}\label{mint}
- m_- \leq m \leq m_+
\end{equation}
with
\begin{equation}\label{m+-}
m_{\pm} = [\sqrt{\beta^2 + \gamma^2} \pm \beta] \,.
\end{equation}
For these partial-waves the parameter $\nu$ is imaginary and the scattering becomes inelastic since $|S_m|\neq 1$.


\subsection{Attraction causes inelastic scattering}

We turn now to a discussion of the physical properties of the modes in the interval $[-m_-,m_+].$ This will at the same time allow us to decide which sign of the square root of $\nu^2$ of Eq.(\ref{nu}) we have to take as index of $J_\nu (\rho).$ Just as in section 1.3 the absorption coefficient is given by $\xi_m=1-|S_m|^2$ whereby $|S_m|^2$ takes now the form
\begin{equation}\label{xi}
|S_m|^2=e^{2\pi\mu} \quad {\rm with} \quad \mu:={\rm Im}\nu = \pm \sqrt{\beta^2+\gamma^2-(m-\beta)^2}, \quad m\in [-m_-, m_+]\,.
\end{equation}
It is clear from the physical point of view that the charged wire will attract the polarizable atoms. This is an effect which may be accompanied by inelastic scattering but in any case not by creation of atoms. Accordingly we must have $\xi_m>0,\, |S_m|^2<1,\,\mu<0$ and therefore the negative sign of the square root in Eq.(\ref{xi}) has to be chosen.

This becomes even more evident when we look at the $\rho$-component of the partial current
\begin{equation}\label{jm}
j_m={\hbar p\over 2iM}\,|c_m|^2\left[J_{-i\mu} (p\rho) J'_{i\mu} (p\rho) - J_{i\mu} (p\rho) J'_{-i\mu} (p\rho)\right] =
{\hbar\over\pi M \rho}e^{\pi\mu}\,{\rm sh}\pi\mu
\end{equation}
with $c_m$ of Eq.(\ref{c}). For the flux to be ingoing we need again $\mu<0.$ Only the modes out of the interval $[-m_-, m_+]$ contribute to this flux and are therefore scattered inelastically. The others for which $\nu^2$ is positive are scattered elastically, if we do not introduce an additional source for absorption, as we will do in the next section. We will attribute to the interval $[m_-, m_+]$ a {\sl critical radius} $\rho_c$ in just the same way as we have related the radius of absorption $\rho_{\rm abs}$ to $[n_-, n_+].$ Classically $\rho_{\rm c}$ represent the impact parameter for which particles spiral down to $\rho\rightarrow 0.$

The total partial flux through a cylindrical surface per length is according to Eq.(\ref{jm})
\begin{equation}\label{jm'}
j^{\rm tot}_m(\rho) = {2\hbar \over M}e^{\pi\mu}\,{\rm sh}\pi\mu
\end{equation}
and the total flux becomes
\begin{equation}\label{j'}
J = {2\hbar\over M}\sum_{m=-m_-}^{m_+}e^{\pi\mu}\,{\rm sh}\pi\mu\,.
\end{equation}
Note that both fluxes are independent of the radius $\rho$ which means that there is the same finite ingoing flux for arbitrarily small values of $\rho.$ This is evidently an unphysical situation. In connection with this problem it has been pointed out in Ref.\cite{Hagen96} that the system, as it is introduced above, does not allow quantum mechanically well-defined solutions. We have therefore to conclude that an extensionless wire - an object which is anyway unphysical - can not even be regarded as an acceptable limit. We will therefore pass on to a different physical problem.

\subsection{Total absorption on the surface of the charged wire}

The physical realistic situation is characterized by a wire of finite radius $\rho_{\rm abs}>0,$ at the surface of which the polarization of the atoms falling in is completely destroyed. This can be represented in our approach in assuming that partial-waves which correspond to the respective impact parameters are totally absorbed. For them we have
\begin{equation}\label{s=0'}
S_m = 0 \quad \mbox{for} \quad -n_- \leq m \leq n_+\,.
\end{equation}
Because of this modification the problem is not further more described in all aspects by the Lagrangian (\ref{L}). We will show below that this absorption induced cutoff will lead to interference effects. The corresponding experiment can be carried out.

The radius of absorption $\rho_{\rm abs}$ may be larger, smaller or equal to the critical radius $\rho_c$ which corresponds to the values of $m$ out of the interval $[-m_-, m_+]$. In the first case, atoms are absorbed which would otherwise be elastically scattered. We have $[-m_-, m_+] \subset [-n_-, n_+]$ and may write the scattering amplitude of Eqs.(\ref{psa'}) and (\ref{phs'}) in the form ($\varphi\neq 0$):
\begin{equation}\label{tf}
f^{\rm abs}(\varphi) = {e^{-i{\pi\over 4}}\over\sqrt{2\pi p}}
\left[\Sigma_1(\beta, \varphi) + \Sigma_2(\beta, \varphi) +
\Sigma_3(\beta, \varphi)\right]
\end{equation}
whereby
\begin{eqnarray}\label{st}
\Sigma_1(\beta, \varphi) &:=& \sum_{m=n_++1}^{\infty}\left(e^{i\pi(m-\nu)}-\cos\pi\beta\right)\, e^{im\varphi}, \nonumber \\
\Sigma_2(\beta, \varphi) &:=& \sum_{m=n_-+1}^{\infty}\left(e^{i\pi(m-\tilde\nu)}-\cos\pi\beta\right)\, e^{-im\varphi}, \nonumber\\
\Sigma_3 (\beta, \varphi) &:=& - \cos\pi\beta\left[\sum_{m=-n_-}^{n_+}\, e^{im\varphi} \right]\,.
\end{eqnarray}
and $\tilde\nu=\nu(-\beta)$.

If on the other hand the radius of absorption is smaller than the critical radius, we have $[-n_-, n_+]\subset [-m_-, m_+].$ Now only the infalling atoms are absorbed on the surface of the wire and the elastically scattered atoms remain undisturbed. In this case the scattering amplitude is of the form Eq.(\ref{tf}) with
\begin{eqnarray}\label{ts}
\Sigma_1(\beta, \varphi) &:=& \sum_{m=n_++1}^{m_+}\left(e^{i\pi m+\pi\mu}-\cos\pi\beta\right)\, e^{im\varphi} \nonumber \\
&+& \sum_{m=m_++1}^{\infty}\left(e^{i\pi(m-\nu)}-\cos\pi\beta\right)\, e^{im\varphi}, \nonumber \\
\Sigma_2(\beta, \varphi) &:=& \sum_{m=n_-+1}^{m_-}\left(e^{i\pi m +\pi\tilde\mu}-\cos\pi\beta\right)\, e^{-im\varphi}\nonumber \\
&+& \sum_{m=m_-+1}^{\infty}\left(e^{i\pi(m-\tilde\nu)}-\cos\pi\beta\right)\, e^{-im\varphi}, \nonumber\\
\Sigma_3 (\beta, \varphi) &:=& - \cos\pi\beta\left[\sum_{m=-n_-}^{n_+}\, e^{im\varphi} \right]
\end{eqnarray}
and $\tilde\mu=\mu(-\beta).$ These expressions are invariant under the substitution $\beta\rightarrow -\beta,\, \varphi\rightarrow -\varphi.$ Therefore we may restrict to $\beta > 0.$ If $m_\pm = n_\pm$ Eqs.(\ref{st}) and (\ref{ts}) agree.

The infinite sums in Eqs.(\ref{st}) and (\ref{ts}) must be given a meaning by means of Eq.(\ref{s+-}) as this was done above for the AB amplitude. Unlike the AB case it is impossible to evaluate the amplitude $f^{\rm abs}(\varphi)$ analytically in a closed expression because the parameter $\nu$ depends nonlinearly on $m.$ To prepare an analytical discussion and numerical calculations we subtract the partial-waves of the AB scattering with absorption from the partial-waves above to obtain more rapidly converging series. Based on this splitting the scattering amplitude (\ref{tf}) may be written in the form
\begin{equation}\label{stf}
f^{\rm abs}(\varphi) = f_{AB}^{\rm abs} (\varphi) + f_w(\varphi)
\end{equation}
where $f_{AB}^{\rm abs} (\varphi)$ is given by the closed expression (\ref{modsa1}). From the physical point of view this has the advantage that the influence of the charged wire is localized in the rest amplitude $f_w(\varphi).$

This rest amplitude may be written in the form
\begin{equation}\label{wf}
f_w (\varphi) = {e^{-i{\pi\over 4}}\over\sqrt{2\pi p}}
\left[\Delta\Sigma_1(\beta, \varphi) + \Delta\Sigma_2(\beta, \varphi)\right]\,.
\end{equation}
For the case $[-m_-, m_+]\subset [-n_-, n_+]$ we find
\begin{eqnarray} \label{dst}
\Delta\Sigma_1(\beta, \varphi) &=& \sum_{m=n_++1}^{\infty}\left(e^{i\pi (m-\nu)}-e^{i\pi(m - |m-\beta|)}\right)\, e^{im\varphi} \nonumber \\
&=& e^{i\pi\beta} \sum_{m=n_++1}^{\infty}\left(e^{i\pi{\beta^2+\gamma^2\over m-\beta+\nu}} - 1 -i{\pi\over 2} {\beta^2+\gamma^2\over m}\right)\, e^{im\varphi} + \delta\Sigma_1(\beta, \varphi) \\
\Delta\Sigma_2(\beta, \varphi) &=& \sum_{m=n_-+1}^{\infty}\left(e^{i\pi (m-\tilde\nu)}-e^{i\pi(m - |m+\beta|)}\right)\, e^{-im\varphi} \nonumber  \\
&=& e^{-i\pi\beta} \sum_{m=n_-+1}^{\infty}\left(e^{i\pi{\beta^2+\gamma^2\over m+\beta+\tilde\nu}} - 1-i{\pi\over 2}{\beta^2+\gamma^2\over m}\right)\, e^{-im\varphi} + \delta\Sigma_2(\beta, \varphi)\nonumber
\end{eqnarray}
with (see the formula 5.4.2(9) in \cite{Prud} )
\begin{eqnarray}\label{dst'}
&&\delta\Sigma_{1, 2}(\beta, \varphi) = i{\pi\over 2} {\beta^2+\gamma^2\over m}\,e^{\pm i\pi\beta}\sum_{m=n_++1}^{\infty}{1\over m}e^{im\varphi} \\
&=& i{\pi\over 2} {\beta^2+\gamma^2\over m}\,e^{\pm i\pi\beta} \left(- \ln (2 \sin{\varphi\over 2}) + i{\pi-\varphi\over 2} - \sum_{m=1}^{n_\pm} {1\over m}\, e^{\pm im\varphi}\right).\nonumber
\end{eqnarray}
For $[-n_-, n_+]\subset [-m_-, m_+]$ we get
\begin{eqnarray} \label{dts}
\Delta\Sigma_1(\beta, \varphi) &=& \sum_{m=n_++1}^{m_+}\left(e^{i\pi m+\pi\mu}-e^{i\pi(m - |m-\beta|)}\right)\, e^{im\varphi} \nonumber \\
&+& \sum_{m=m_++1}^{\infty}\left(e^{i\pi (m-\nu)}-e^{i\pi(m - |m-\beta|)}\right)\, e^{im\varphi} \nonumber \\
&=& \sum_{m=n_++1}^{m_+}\left(e^{i\pi m+\pi\mu}-e^{i\pi(m - |m-\beta|)}\right)\, e^{im\varphi} \nonumber \\
&+& e^{i\pi\beta} \sum_{m=m_++1}^{\infty} \left(e^{i\pi{\beta^2+\gamma^2\over m-\beta+\nu}} - 1 -i{\pi\over 2} {\beta^2+\gamma^2\over m}\right)\, e^{im\varphi}+ \delta\Sigma'_1(\beta, \varphi)\,, \\
\Delta\Sigma_2(\beta, \varphi)&=& \sum_{m=n_-+1}^{m_-}\left(e^{i\pi m+\pi\tilde\mu}-e^{i\pi(m - |m+\beta|)}\right)\, e^{-im\varphi} \nonumber \\
&+& \sum_{m=m_-+1}^{\infty}\left(e^{i\pi (m-\tilde\nu)}-e^{i\pi(m - |m+\beta|)}\right)\, e^{-im\varphi} \nonumber \\
&=& \sum_{m=n_-+1}^{m_-}\left(e^{i\pi m+\pi\tilde\mu}-e^{i\pi(m - |m+\beta|)}\right)\, e^{-im\varphi} \nonumber \\
&+& e^{-i\pi\beta} \sum_{m=m_-+1}^{\infty} \left(e^{i{\pi\over 2}{\beta^2+\gamma^2\over m+\beta+\tilde\nu}} -1 -i{\pi\over 2} {\beta^2+\gamma^2\over m}\right)\, e^{-im\varphi}+ \delta\Sigma'_2(\beta, \varphi)\nonumber
\end{eqnarray}
with $\delta\Sigma'_{1, 2}(\beta, \varphi)=\delta\Sigma_{1, 2}(\beta, \varphi)|_{n_\pm\rightarrow m_\pm}.$ We subtracted the sums $\delta\Sigma_{1, 2}(\beta, \varphi)$ from $\Delta\Sigma'_{1, 2}(\beta, \varphi)$ to get uniformly converging series in the rest. For $n_\pm = m_\pm$ this is result we have obtained in \cite{Audretsch98}.

Now the behaviour of the scattering amplitude (\ref{stf}) for small scattering angles can be discussed analytically. The main contribution arises from the modified AB amplitude (\ref{modsa1})
\begin{equation}\label{singAB}
f_{AB}^{\rm abs}(\varphi\ll 1) = - {e^{-i{\pi\over 4}}\over\sqrt{2\pi p}}\,e^{i{n_+-n_-\over 2}\varphi}\, {\sin\pi\beta\over \sin {\varphi\over 2}}\,\cos {n^{\rm abs}\varphi\over 2}.
\end{equation}
We keep the factor $\cos n^{\rm abs}\varphi/2$ which leads for large $n^{\rm abs}$ to deviations from the background of the singular AB term. The remaining part of the modified AB amplitude $f_{AB}^{\rm abs}(\varphi)$ of Eq.(\ref{modsa1}) is regular for $\varphi=0.$ Other, less singular terms are contained in terms $\delta\Sigma_{1, 2}(\beta, \varphi).$ They arise due to the interaction of the polarizable atoms with the charge wire.  Neglecting small terms in Eq.(\ref{wf}) we obtain
\begin{equation}\label{singw}
f_w(\varphi \ll 1) \rightarrow -i \pi (\beta^2 + \gamma^2) \cos\pi\beta \ln (2 \sin{\varphi\over 2})\,.
\end{equation}
We note the fact that the presence of the magnetic field suppresses this singularity because of the term $\cos\pi\beta.$.

In the presence of the magnetic field $(\beta\neq 0)$ the AB effect with absorption dominates for the small angles $\varphi\rightarrow 0$ the influence of the wire according to
$$
{f_w\over f_{AB}^{\rm mod}} \sim \cos\pi\beta\,\varphi\ln\varphi\rightarrow 0\,.
$$
The appearance the AB term for the non-vanishing magnetic field should therefore be observable in any case. If the magnetic field is absent the term (\ref{singw}) gives the main contribution to the scattering for small angles.

Since the total absorption cuts out the partial modes in some interval of $m,$ the scattering amplitude has, as in the corresponding AB case, an oscillatory behaviour. These oscillation has the width of order $2\pi/n^{\rm abs}$ and are relevant for angles $\varphi \sim 2\pi/n^{\rm abs}.$ If the angular resolution is good enough they can be observed. If not, the cross section will be averaged leading to a damping factor 1/2.
\begin{figure}[h]
\let\picnaturalsize=N
\def\picsize{2.5in}
\ifx\nopictures Y\else{\ifx\epsfloaded Y\else\input epsf \fi
\let\epsfloaded=Y
{\hspace*{\fill}
\parbox{2.5in}{\ifx\picnaturalsize N\epsfxsize \picsize\fi \epsfbox{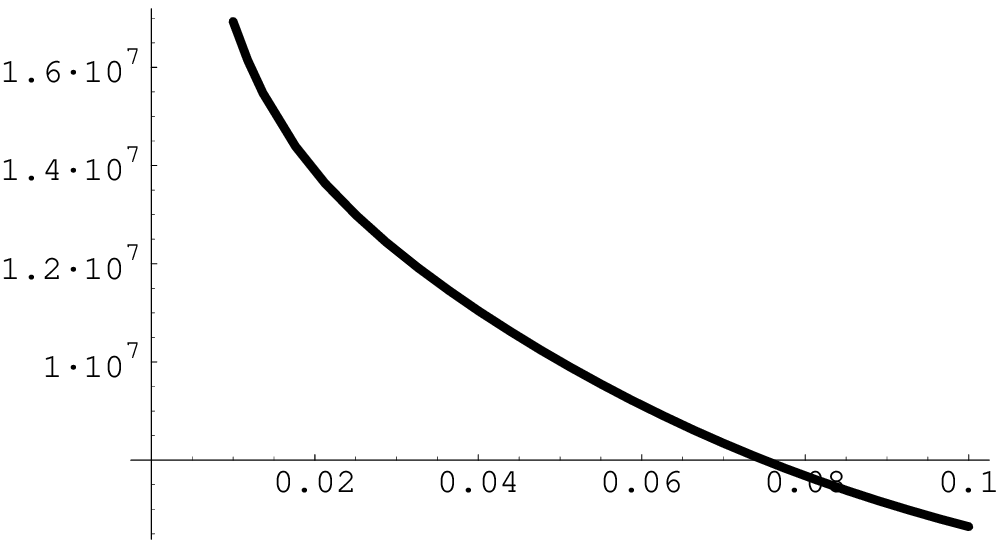}}\hfill
\put(2,-32){{\small $\varphi$}}
\put(-154,56){{\small $\eta$}}
\put(210,-44){{\small $\varphi$}}
\put(44,62){{\small $\eta$}}
\put(-115,16){{\tiny $\leftarrow\beta=0.5$}}
\put(90,16){{\tiny $\leftarrow\beta=0.5$}}
\hspace{0.8cm}
\parbox{2.5in}{\ifx\picnaturalsize N\epsfxsize \picsize\fi \epsfbox{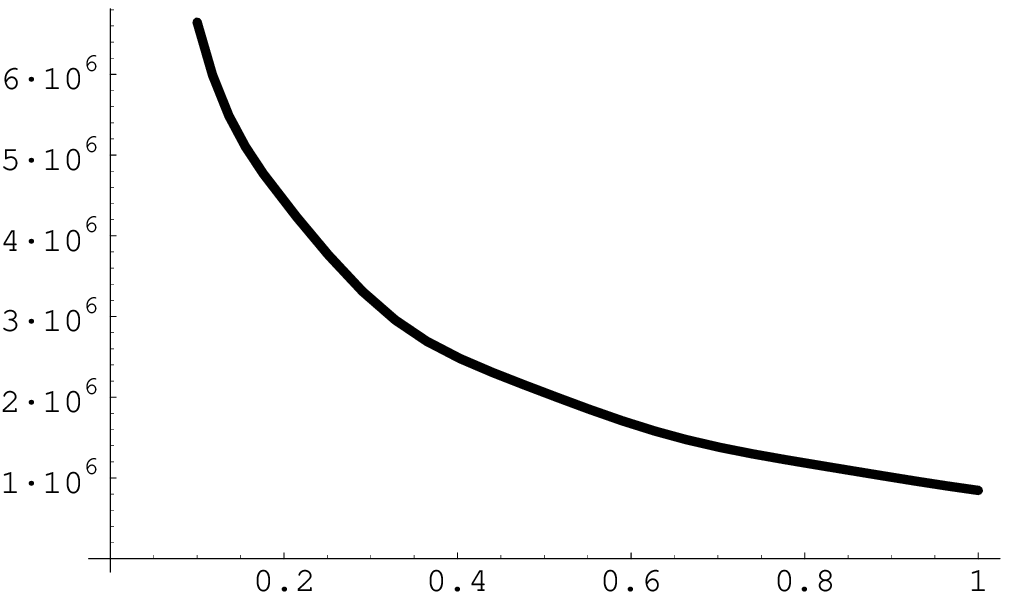}}\hspace*{\fill}}}\fi\\
\par\vspace{0.5cm}

\let\picnaturalsize=N
\def\picsize{2.5in}
\ifx\nopictures Y\else{\ifx\epsfloaded Y\else\input epsf \fi
\let\epsfloaded=Y
{\hspace*{\fill}
\parbox{2.5in}{\ifx\picnaturalsize N\epsfxsize \picsize\fi \epsfbox{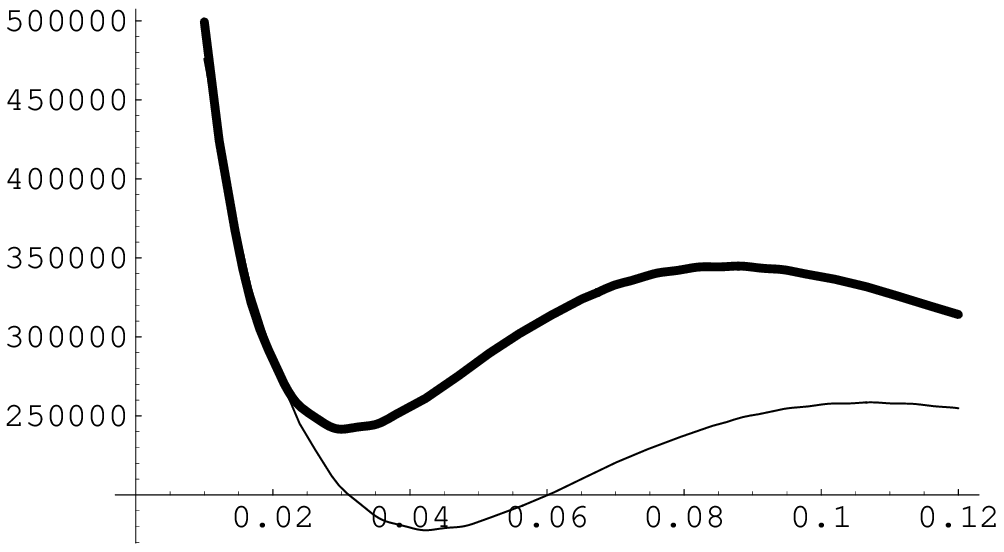}}\hfill
\put(10,-40){{\small $\varphi$}}
\put(-158,58){{\small $\eta$}}
\put(204,-34){{\small $\varphi$}}
\put(37,60){{\small $\eta$}}
\put(-115,0){{\tiny $\beta=0.1 \rightarrow$}}
\put(-100,-30){{\tiny $\beta=0 \rightarrow$}}
\put(110,0){{\tiny $\leftarrow\beta=0.1$}}
\put(75,-20){{\tiny $\beta=0 \rightarrow$}}
\hspace{0.5cm}
\parbox{2.5in}{\ifx\picnaturalsize N\epsfxsize \picsize\fi \epsfbox{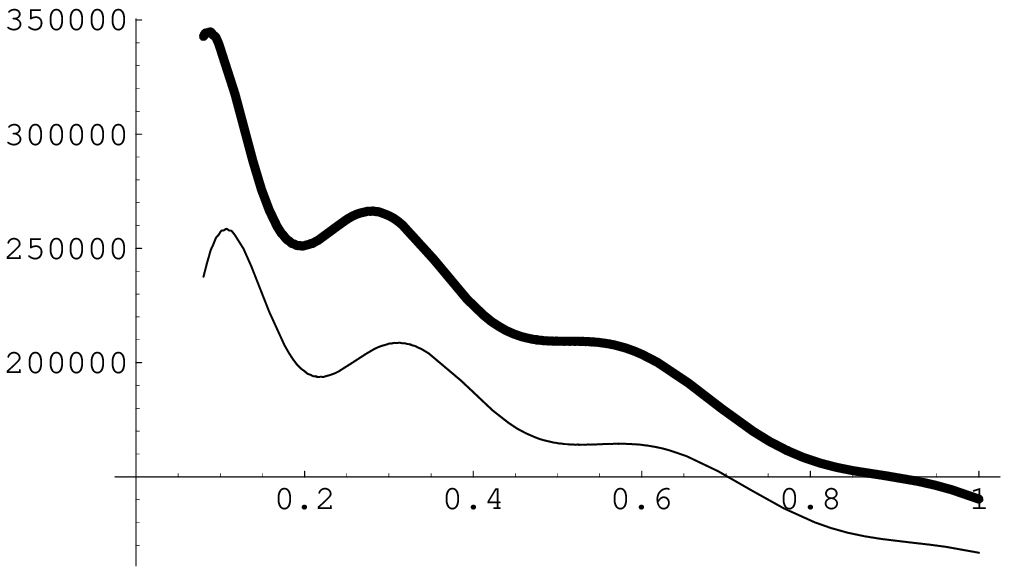}}\hspace*{\fill}}}\fi\\
\par\vspace{0.5cm}
\caption{Dimensionless differential cross section $\eta = 2\pi p|f^{\rm abs}(\varphi)|^2$ for the elastic scattering of neutral atoms as function of the scattering angle $\varphi$ for the different values of the magnetic field parameter $\beta.$ It is assumed that the radius of absorption is larger that the critical radius. The number of totally absorbed modes is $n^{\rm abs}=20.$} For $\beta=0$ there is only a charged wire with a totally absorbing surface (thin line).
\label{fig-label}
\end{figure}

The behaviour of the scattering cross section for arbitrary angles can be found by numerical calculation and is shown in Fig.2 which refers to the case that the radius of absorption is larger than the critical radius. If we compare these figures with Fig.1, we see that the AB effect for atoms leads for the same number of absorbed modes and the same magnetic flux parameter $(\beta=0.5)$ to a considerably larger cross section as in the case of the charged particles. For vanishing magnetic field ($\beta=0,$ i.e. only a charged wire with absorption on its surface is present) the cross section is already of this order and it increases with $\beta.$ The characteristic oscillations are already there for $\beta=0.$ They remain for increasing magnetic flux parameter $\beta$ and practically disappear for $\beta=0.5.$ If a larger number of absorbed modes is taken, the cross section of the elastic scattering, the amplitude and frequency of the oscillations all go up.

\subsection{Limiting cases}

If one wants to leave scruples apart in assuming that the wire has a vanishing radius, there are no partial-waves with vanishing $S_m$ in this case. This situation can be easily obtained as a special case of Eqs.(\ref{dts}) and (\ref{modsa}) with $n_+=N,\;n_-=-N-1.$ The scattering amplitude reads
\begin{equation}\label{stf'}
f(\varphi) = f_{AB} (\varphi) + {\tilde f_w}(\varphi)
\end{equation}
with the unmodified AB amplitude $f_{AB}$ of Eq.(\ref{sa}) and the rest amplitude
\begin{equation}\label{wf'}
{\tilde f_w}(\varphi) = {e^{-i{\pi\over 4}}\over\sqrt{2\pi p}}
\left[\Delta{\tilde\Sigma_1}(\beta, \varphi) + \Delta{\tilde\Sigma_2}(\beta, \varphi)\right]
\end{equation}
whereby
\begin{eqnarray} \label{dts'}
\Delta{\tilde\Sigma_1}(\beta, \varphi) &=& \sum_{m=1}^{m_+}\left(e^{i\pi m+\pi\mu}-e^{i\pi(m - |m-\beta|)}\right)\, e^{im\varphi} \nonumber \\
&+&\sum_{m=m_++1}^{\infty}\left(e^{i\pi (m-\nu)}-e^{i\pi(m - |m-\beta|)}\right)\, e^{im\varphi} \nonumber \\
&=& \sum_{m=1}^{m_+}\left(e^{i\pi m+\pi\mu}-e^{i\pi(m - |m-\beta|)}\right)\, e^{im\varphi} \nonumber \\
&+& e^{i\pi\beta} \sum_{m=m_++1}^{\infty} \left(e^{i\pi {\beta^2+\gamma^2\over m-\beta+\nu}}-1 -i{\pi\over 2} {\beta^2+\gamma^2\over m}\right) e^{im\varphi}+ \delta\Sigma'_1(\beta, \varphi)\,, \nonumber\\
\Delta{\tilde\Sigma_2}(\beta, \varphi)&=& \sum_{m=0}^{m_-}\left(e^{i\pi m+\pi\tilde\mu}-e^{i\pi(m - |m+\beta|)}\right)\, e^{-im\varphi} \\
&+& \sum_{m=m_-+1}^{\infty}\left(e^{i\pi (m-\tilde\nu)}-e^{i\pi(m - |m+\beta|)}\right)\, e^{-im\varphi} \nonumber \\
&=& \sum_{m=0}^{m_-}\left(e^{i\pi m+\pi\tilde\mu}-e^{i\pi(m - |m+\beta|)}\right)\, e^{-im\varphi} \nonumber \\
&+& e^{-i\pi\beta} \sum_{m=m_-+1}^{\infty}\left(e^{i\pi {\beta^2+\gamma^2\over m+\beta+\tilde\nu}} - 1 -i{\pi\over 2} {\beta^2+\gamma^2\over m}\right)e^{-im\varphi}+ \delta\Sigma'_2(\beta, \varphi)\,,\,.\nonumber
\end{eqnarray}

If on the other hand there is only a charged wire and no magnetic field at all ($\beta=0$), the  scattering amplitude contains no AB contribution and reduces in the case of vanishing wire radius to
\begin{equation}\label{wf''}
f(\varphi)) = {e^{-i{\pi\over 4}}\over\sqrt{2\pi p}}
2\left[ \sum_{m=0}^{[\gamma]}\left(e^{i\pi m+\pi\mu}-1\right)\, \cos m \varphi + \sum_{m=[\gamma]+1}^{\infty}\left(e^{i\pi (m-\nu)}- 1\right)\, \cos m \varphi\right]. \nonumber
\end{equation}

\section{Conclusion}

If taken with some care, the partial-wave analysis of elastic and inelastic scattering can successfully be applied to situations where an AB potential is present. We have done this for charged particles moving in an AB potential when absorption is present, and for neutral polarizable atoms moving in the field of an electrically charged wire aligned to a homogeneous magnetic field. Atoms colliding with the wire undergo inelastic scattering. The case of total absorption on the wire surface is discussed for different choices of the relevant parameters. Absorption leads in all cases to characteristic oscillations of the cross section with the scattering angle. The AB effect with absorption dominates the scattering near the forward direction.

\bigskip
{\bf Acknowledgments}
\medskip

V.~S.~thanks Prof.J.~Audretsch and the members of his group for the
friendly atmosphere and the hospitality at the University of Konstanz.
This work was supported by the Deutsche Forschungsgemeinschaft.


\end{document}